# Filtro Adaptativo y Módulo de Grabación en Dispositivo Para Mejora en la Calidad de Audición

Carlos Elihu Palomino Torres, Francisco Claudio Chichipe Mondragón, Frank Antonio Siesquen Rodriguez, Mariana Alexandra Huaynate León

*Resumen- Este proyecto propone el desarrollo de un sistema para mejorar la percepción auditiva en tiempo real, empleando un ESP32, un filtro adaptativo LMS y técnicas de inteligencia artificial. El sonido es captado mediante un micrófono I2S INMP44, procesado dinámicamente para minimizar el ruido y luego reproducido a través de un speaker MAX98357. Gracias a la adaptación automática del filtrado, el dispositivo es capaz de ajustarse a diversas condiciones acústicas, ofreciendo una experiencia auditiva más clara y eficiente.*

*Palabras clave: ESP32, filtro adaptativo, inteligencia artificial, procesamiento de señales, mejora auditiva.*

*Abstract- This project presents the development of a real-time auditory enhancement system utilizing an ESP32, an LMS adaptive filter, and artificial intelligence techniques. An I2S INMP44 microphone captures the sound, which is dynamically processed to suppress noise before being played through a MAX98357 speaker. The system continuously adapts to varying acoustic environments, ensuring improved speech clarity and an optimized listening experience.*

*Keywords: ESP32, adaptive filter, artificial intelligence, signal processing, auditory enhancement.*

## I. INTRODUCCIÓN

La capacidad de oír con claridad es esencial para la comunicación y la interacción con el entorno. Sin embargo, el ruido ambiental y ciertos problemas auditivos pueden dificultar esta función, generando incomodidad y limitaciones en la vida diaria. Para mitigar estos efectos, el procesamiento digital de señales ha permitido el desarrollo de tecnologías que mejoran la inteligibilidad del sonido en distintas situaciones.

El uso de filtros adaptativos, en particular aquellos basados en el algoritmo LMS, ha demostrado ser una estrategia eficaz para reducir el ruido de fondo sin distorsionar las señales útiles. Además, la inteligencia artificial ha abierto nuevas posibilidades para que estos sistemas se ajusten de manera autónoma a cada entorno y usuario.

Este trabajo presenta la implementación de un dispositivo de asistencia auditiva portátil, basado en un ESP32, que integra un filtro LMS con un modelo de IA para la optimización continua del procesamiento de audio. Se han realizado pruebas en distintos escenarios, evaluando su rendimiento en términos de reducción de interferencias y mejora de la claridad del sonido emitido.

## II. INFORMACIÓN DEL PROYECTO

*A. Descripción:*

El sistema desarrollado está diseñado para mejorar la percepción auditiva en tiempo real mediante procesamiento digital de señales. La arquitectura del dispositivo se basa en un ESP32, un microcontrolador con capacidad de cómputo suficiente para ejecutar algoritmos de filtrado adaptativo sin comprometer la latencia del sistema.

La adquisición del audio se realiza a través de un micrófono I2S INMP44, el cual entrega una señal digital con un alto rango dinámico y una baja relación señal-ruido (SNR), lo que permite capturar con precisión el sonido del entorno. Esta señal es procesada mediante un filtro adaptativo LMS (Least Mean Squares), cuyo coeficiente de actualización se ajusta dinámicamente para minimizar la interferencia de ruido y preservar la inteligibilidad del habla.

Para optimizar la convergencia del filtro y evitar inestabilidades numéricas, el sistema implementa un escalado adaptativo de la señal de entrada y una regularización de los coeficientes del filtro.

El audio procesado es reproducido a través de un módulo de amplificación y salida MAX98357, que opera con modulación PWM (Pulse Width Modulation) optimizada para mantener una baja distorsión armónica (THD) y una alta eficiencia energética. Gracias a la integración de un pipeline de procesamiento optimizado, el sistema logra mantener una latencia mínima, asegurando una respuesta inmediata en la adaptación del filtrado a cambios en el entorno acústico.

*B. Justificación:*

Este proyecto busca mejorar la experiencia auditiva mediante la implementación de un filtro adaptativo basado en inteligencia artificial, capaz de reducir el ruido y optimizar la señal de interés en tiempo real.

El desarrollo de tecnologías accesibles y eficientes en el procesamiento de audio es crucial para crear soluciones asequibles y adaptables a diferentes necesidades. La combinación de filtros adaptativos con aprendizaje automático permite que el sistema se ajuste de manera inteligente a distintos entornos, ofreciendo una mejora sustancial en la percepción del sonido sin necesidad de intervención manual.

Además, este proyecto representa una contribución al avance en el campo del procesamiento de señales en sistemas embebidos, con aplicaciones potenciales en dispositivos de asistencia auditiva, reducción de ruido y comunicación en entornos adversos. Su impacto va más allá de la innovación tecnológica, proporcionando una herramienta que puede mejorar la calidad de vida de muchas personas.

*C. Objetivos:*

Desarrollar un sistema de mejora en la calidad de audición utilizando un filtro adaptativo basado en el algoritmo LMS, implementado en un dispositivo con ESP32, para reducir el ruido y optimizar la percepción del sonido en tiempo real.

Diseñar un sistema escalable que pueda ser de beneficio para personas con alguna discapacidad auditiva, priorizando la transmisión de audio limpia con una latencia mínima.

Escoger el algoritmo de procesamiento adecuado para garantizar una transmisión de datos constante y sin pérdidas significativas, así mismo establecer la configuración y establecimiento adecuado de parámetros para el correcto funcionamiento del sistema.

*D. Materiales:*

- Microcontrolador ESP32
- Modulo I2S INMP44
- Modulo MAX98357
- Módulo ISD1820
- Interfaz ARDUINO IDE
- Pc's

## III. DESARROLLO DE CONTENIDOS

*A. Filtros Adaptativos:*

El filtro adaptativo es formalmente definido como un dispositivo auto diseñado con parámetros de variación en el tiempo que son ajustados continuamente de acuerdo con los valores de entrada. En la realidad, el filtro adaptativo es no lineal dado que no obedece al principio de superposición.

Se puede clasificar el filtro adaptativo de acuerdo con las características de interés.

*B. Filtros Adaptativos Lineales y No Lineales*

A pesar del comportamiento no lineal inherente de los filtros adaptativos, estos son clasificados en filtros adaptativos lineales o no lineales dependiendo si las unidades computacionales básicas en su construcción son lineales o no. Específicamente, para un filtro adaptativo lineal, la estimación de la cantidad de interés es calculada en la salida del filtro como una combinación lineal del conjunto disponible de observaciones aplicadas en la entrada del filtro. un filtro adaptativo lineal incluye una unidad única computacional por cada salida. Ejemplos de este tipo de filtros son el algoritmo LMS y el RLS.

En contraparte, los filtros adaptativos no lineales incluyen el uso de elementos computacionales no lineales que hacen posible explorar la información contenida en los datos de entrada, su naturaleza dificulta el análisis matemático de su comportamiento en contraste con los filtros adaptativos lineales. Ejemplos significativos de este tipo de filtros incluye a los filtros de Volterrra y las redes neuronales.

*C. Filtros Adaptativos Recursivos y No Recursivos*

Esta clasificación se basa en términos de si la construcción física del filtro incluye o no cualquier forma de feedback. Un filtro adaptativo no recursivo tiene una memoria finita mientras que el recursivo tiene una memoria infinita que decae con el tiempo. Un ejemplo de el filtro adaptativo no recursivo es tapped-delay-line filter mientras que la respuesta al impulso infinito es un ejemplo de un filtro recursivo.

Es importante remarcar que el término "recursivo" no se refiere al ajuste algorítmico de los parámetros del filtro, más bien, como se mencionó con anterioridad, a la presencia de alguna forma de realimentación en la construcción física del filtro.

*D. Filtros Adaptativos Supervisados y No Supervisados.*

Esta clasificación está basada en como la respuesta deseada es provista. En caso de un filtro adaptativo supervisado este requiere de un maestro para proporcionar la respuesta deseada.

Del otro lado, se hablará de filtros adaptativos no supervisado en el que el proceso de aprendizaje de corrección de errores se desarrolla sin la necesidad de una aportación independiente (es decir, el profesor) que proporcione la respuesta deseada.

*E. Algoritmos LMS:*

El acercamiento al algoritmo Least Mean Square se muestra en la Fig. donde $X_k = [x_{0k}, x_{1k} \ldots x_{N-1k}]^T$ es la entrada de los vectores de datos en el instante $kth$, $y_k$ es la señal deseada y $\hat{y}_k$ representa su estimación. La señal puede ser estimada correctamente por el filtro con valores adecuados de su coeficiente $W_k$, el cual es obtenido a través de la minimización del cuadrado de la señal de error $e_k$. Entonces, mientras un sistema aprende de su ambiente, este es diseñado como un filtro adaptativo donde los coeficientes del filtro son ajustados de forma recursiva hacia sus valores óptimos.

Para cada iteración, el peso de cada vector es obtenido como.

$$W_{K+1} = W_K + \mu(-\nabla_k)$$

Donde $\mu$ es el parámetro de adaptación, $W_k$ es $[w_{0k}, w_{1k} \ldots w_{N-1k}]^T$ los coeficientes del filtro, y $\nabla_k$ es la gradiente de la superficie de rendimiento de error con respecto a el coeficiente del filtro $W_k$ el cual puede ser encontrado como:

$$\widehat{\nabla}_k = -2e_k X_k$$

La recursión [1] se denomina algoritmo LMS y es inicializado cuando se establece todos los coeficientes en cero. Después, el algoritmo procede al cálculo de la señal de error $e_k$ que se utiliza para calcular los coeficientes actualizados. Este procedimiento se realiza hasta que se alcancen las condiciones de un estado estable. La estabilidad de un sistema de bucle cerrado de este tipo está regida por el parámetro de adaptación $\mu$ y debe satisfacer la siguiente condición:

$$0 < \mu < \frac{2}{total\ input\ power}$$

Donde la potencia de entrada se refiere a la suma de los valores de los cuadrados medios de las entradas. Cuando $\mu$ es pequeño, el algoritmo LMS toma mas tiempo para aprender sobre su entrada con el mínimo error del cuadrado medio y viceversa. Por lo tanto, se prefiere una secuencia de pasos de tamaño variable en el tiempo para una mejor convergencia.

*F.     Algoritmos SLMS*

Los algoritmos SLMS identifican el algoritmo LMS inicial, y reducen el número de operaciones matemáticas realizadas, dependiendo de los signos de los datos de entrada o del signo del error. Considere el signo de error, se altera la magnitud de la corrección en la actualización del vector de coeficientes, mientras que la dirección, gradiente estimado, es compatible con el algoritmo LMS equivalente con una convergencia inversamente proporcional a la magnitud del error. Otra opción para los algoritmos signados es el algoritmo LMS con el signo del dato que altera la dirección del vector actualizado. Por esta razón, es menos confiable que LMS, y en algunos casos los coeficientes divergen mientras alcanzan la convergencia.[2]

*G.     Protocolo I2S*

Estándar eléctrico de bus serial usado para interconectar circuitos de audio digital. El I2S separa las señales de datos y de reloj, lo que resulta en menores cantidad de fluctuación de la señal que en sistemas que recuperan el reloj de la señal de datos[3].

Puede manejar únicamente datos de audio. Contando con 3 líneas, estas consisten en una línea de dos canales multiplexados (SD) un Word select (WS) y un clock (SCK)

Para la transmisión de datos, tanto el transmisor como el receptor utilizas la misma señal de reloj. En un sistema básico, el transmisor se encarga de generar la señal SCK, la señal WS y la SD. Sin embargo, en sistemas complejo, donde hay múltiples transmisores y receptores, se emplea un dispositivo específico que asume el rol de maestro y gestiona el flujo de datos entre los diferentes dispositivos.

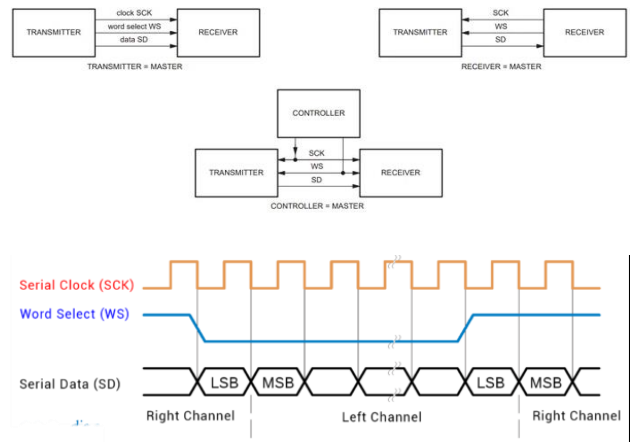

*Fig. 01 - Diagrama de tiempos I2S*

*H.     INMP441*

Micrófono omnidireccional de alto rendimiento, bajo consumo de energía y salida digital con un puerto inferior. Cuenta con un acondicionamiento de señal, un convertidor analógico digital, administración de energía y una interfaz I2S de 24bits.

El micrófono MEMS detecta ondas de presión sonora y las convierte en una señal eléctrica interna, esta se digitaliza internamente mediante un modulador, la señal digitalizada se envía a al microcontrolador a través de la interfaz I2S.[4]

*I.     MAX98357A*

I2S - MAX98357A es una herramienta potente que emplea el convertidor digital a analógico (DAC) MAX98357A para convertir señales de audio en formato I2S en una salida analógica capaz de alimentar altavoces. Esta placa puede suministrar hasta 3.2 W de potencia a una carga de 4 Ω y permite configurarse para reproducir audio en el canal izquierdo, derecho o en ambos.[5]

La entrada de audio es estándar I2S, puede usar datos lógicos de 3.3V o 5V. Las salidas son "Bridge Tied" eso significa que se conectan directamente a las salidas, sin conexión a tierra. La salida es un PWM de onda cuadrada de 300 KHz que luego es 'promediado' por la bobina del altavoz; las altas frecuencias no se escuchan. Todo lo anterior significa que no puede conectar la salida a otro amplificador, debería impulsar los altavoces directamente.

Hay un pin de ganancia que se puede manipular para cambiar la ganancia. De forma predeterminada, el amplificador le dará 9 dB de ganancia. Conectando una resistencia pullup o pull down, o cableando directamente, el pin de ganancia se puede configurar para dar 3dB, 6dB, 9dB, 12dB o 15dB.

El pin ShutDown / Mode se puede usar para apagar el chip o configurar qué canal de audio I2S se conecta al altavoz. De forma predeterminada, el amplificador emitirá (L + R) / 2 mezcla estéreo en salida mono. Al agregar una resistencia, puede cambiarla para que sea solo la salida izquierda o derecha.

### J. Grabador de Voz y Parlante ISD1820

ISD1800 grabador y reproductor de un solo chip y de un solo mensajes. Con duración seleccionable por el usuario de 6 a 16 segundos. Los dispositivos CMOS incluyen chip oscilador, preamplificador de micrófono, control automático de ganancia, filtro antialiasing, matriz de almacenamiento multinivel, filtro suavizante y amplificador de altavoz. El subsistema se puede configurar con un micrófono, un altavoz, varios componentes pasivos, dos pulsadores, botones y una fuente de alimentación. Las grabaciones se almacenan en celdas de memoria no volátil en el chip, lo que proporciona almacenamiento de mensajes de energía cero.

## IV. DESARROLLO DEL PROYECTO

### A. Diseño Conceptual:

Basado en el procesamiento digital de señales mediante un filtro adaptativo LMS, con el objetivo de optimizar la percepción del sonido en tiempo real. La arquitectura del dispositivo emplea un microcontrolador ESP32, encargado de gestionar la adquisición, procesamiento y salida del audio. La señal es capturada por un micrófono digital I2S INMP44, procesada para minimizar el ruido ambiental mediante el algoritmo LMS y luego amplificada con un módulo MAX98357 para su reproducción. Este enfoque permite que el sistema se adapte dinámicamente a diferentes entornos acústicos, mejorando la claridad del sonido sin necesidad de intervención manual.

En la Fig. 02 se presenta el diagrama de bloques de la idea preliminar para suprimir el ruido en las señales de audio.

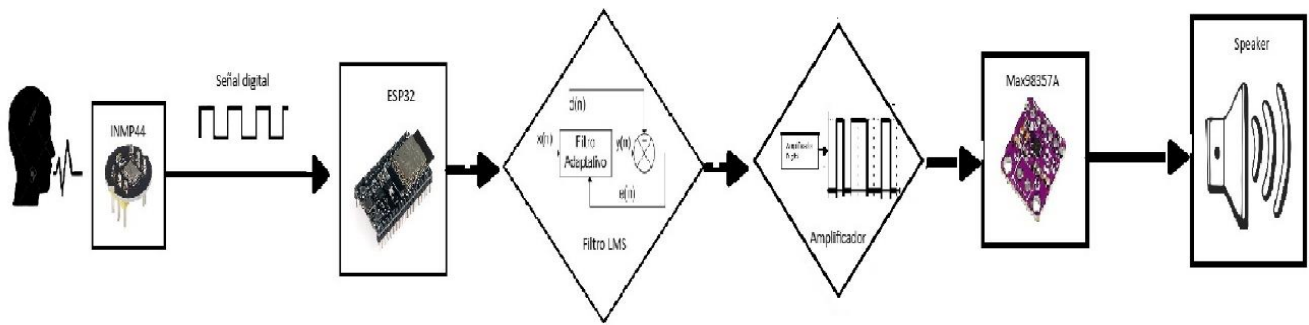

*Fig. 02. - Diagrama de bloques del funcionamiento del Filtro Adaptativo y Módulo de Grabación*

### B. Diseño Estructural:

Se explican a continuación las etapas que estarán presentes en el diseño.

#### 1) Captura de la Señal de Audio:

Se inicia con la captura del audio ambiental. Esto se logra mediante el micrófono digital I2S INMP44, un transductor que convierte las ondas sonoras en señales eléctricas de alta fidelidad. Este micrófono está basado en una interfaz I2S (Inter-IC Sound), un protocolo ampliamente utilizado para la transmisión digital de audio, que evita pérdidas de calidad típicas en conversiones analógicas a digitales [6].

El INMP44 proporciona una señal digitalizada con una alta relación señal-ruido (SNR) y baja distorsión, lo cual es crucial para mantener la claridad del sonido original [7].

Además, su frecuencia de muestreo de hasta 48 kHz es ideal para capturar el rango completo de frecuencias auditivas, lo que asegura una representación precisa del sonido [8]. Esta señal digital se envía directamente al ESP32 a través de la interfaz I2S, garantizando una transmisión eficiente sin latencia significativa [9].

El uso de micrófonos MEMS, como el INMP44, ha demostrado ser eficaz en aplicaciones de procesamiento de audio en tiempo real, especialmente en sistemas embebidos de bajo consumo [9], [10]

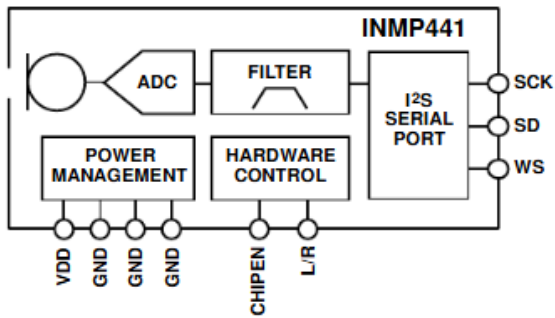

*Fig. 03 - Diagrama de bloques funcional*

2) *Reprocesamiento de la Señal:*

Una vez que la señal digital es capturada, pasa a la etapa de preprocesamiento. En esta fase, se aplican técnicas de normalización y escalado para asegurar que el rango dinámico de la señal sea adecuado para el posterior procesamiento. Esto se logra utilizando un algoritmo de normalización adaptativa, que ajusta la amplitud de la señal de entrada para evitar la saturación y la distorsión, asegurando que la señal se mantenga dentro de los límites óptimos para el filtrado [7].

Además, se implementa un filtro pasa-bajo preliminar que atenúa las frecuencias no deseadas fuera del rango auditivo de interés, generalmente de 20 Hz a 20 kHz, permitiendo un procesamiento más eficiente [10].

Este filtro también ayuda a reducir el ruido de alta frecuencia y las interferencias, lo cual es crucial para el siguiente paso del procesamiento [11].

La combinación de estas técnicas garantiza que la señal esté en condiciones óptimas para ser procesada por el filtro adaptativo LMS, mejorando así la eficacia del sistema en la cancelación de ruido y la preservación de la señal útil [12].

3) *Filtro Adaptativo LMS:*

El núcleo del procesamiento de señales en este sistema es el filtro adaptativo LMS, que se encarga de minimizar el ruido y las interferencias mientras preserva las señales de interés, como el habla. El algoritmo LMS es un algoritmo de adaptación de mínimos cuadrados, que ajusta los coeficientes del filtro en función de las variaciones en la señal de entrada [13]. El filtro LMS se configura de la siguiente manera: [14].

- Señal de referencia: Se utiliza la señal capturada por el micrófono como referencia para la estimación de la señal de ruido.

- Señal de error: La señal de error es la diferencia entre la señal deseada (idealmente sin ruido) y la salida del filtro. Este error se utiliza para actualizar los coeficientes del filtro de manera iterativa.

- Actualización de coeficientes: Los coeficientes del filtro son actualizados en cada iteración utilizando el algoritmo de descenso del gradiente, lo que garantiza que el filtro se adapte a las condiciones acústicas cambiantes. El factor de adaptación se ajusta cuidadosamente para evitar sobreajuste y asegurar la estabilidad del sistema.

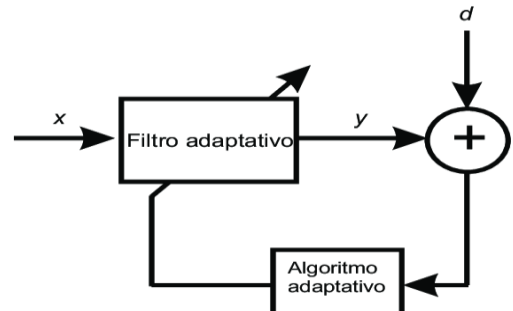

*Fig.04 - Esquema de un filtro adaptativo LMS*

4) *Amplificación y Salida de la Señal:*

Una vez que el audio ha sido procesado y mejorado, la señal es enviada al amplificador MAX98357. Este componente amplifica la señal procesada y la convierte en una salida de audio audible. El MAX98357 es un amplificador de clase D, que proporciona una alta eficiencia energética y baja distorsión armónica total (THD), asegurando que la señal de salida sea lo más fiel posible a la señal procesada sin agregar artefactos o ruidos adicionales [15].

El amplificador es controlado por el ESP32, el cual gestiona la ganancia de salida y la modulación por ancho de pulso (PWM), optimizando la potencia de salida en función de las condiciones del sistema y la demanda energética [16].

La señal amplificada se envía al altavoz, proporcionando una salida clara y nítida, ideal para audífonos o dispositivos portátiles [17].

5) *Grabación de Audio:*

El módulo ISD1820 permite almacenar mensajes de voz en su memoria interna no volátil. La grabación se activa al habilitar la señal en el pin REC, lo que permite almacenar el sonido en la matriz de almacenamiento multinivel del módulo [18].

Durante esta fase, el ISD1820 utiliza su preamplificador interno y control automático de ganancia (AGC) para optimizar la calidad del audio grabado [19].

Una vez finalizada la grabación, el mensaje queda almacenado en el chip sin necesidad de energía adicional, permitiendo su reproducción posterior mediante los pines de control PLAYE o PLAYL, asegurando una respuesta eficiente y una calidad de audio adecuada para su salida a través del altavoz [19].

*C. Programación de la Placa:*

*1) Librerías:*

Para la implementación del sistema, se emplean varias librerías clave. La librería Arduino.h permite la integración con el entorno de desarrollo de Arduino, simplificando la gestión de pines y periféricos, así como el acceso a funciones básicas como digitalWrite y analogRead [20].

Por otro lado, la librería driver/i2s.h es esencial para configurar y controlar el protocolo I2S (Inter-IC Sound), utilizado tanto en la captura de la señal desde el micrófono INMP44 como en la reproducción del audio procesado a través del amplificador MAX98357A [7].

Esta librería facilita la comunicación digital de alta fidelidad, evitando pérdidas de calidad en la transmisión de audio [21].

Además, se incluye la librería math.h, indispensable para realizar operaciones matemáticas avanzadas, como la implementación del filtro LMS (Least Mean Squares) y la normalización de la señal, asegurando que los cálculos sean precisos y eficientes [13].

*2) Parámetros:*

El sistema define varios parámetros críticos para asegurar un procesamiento adecuado del audio. La frecuencia de muestreo se fija en 48 kHz, lo que permite capturar con precisión el rango de frecuencias audibles (20 Hz a 20 kHz) y cumple con estándares de calidad de audio [8]. Se utiliza un búfer de 512 muestras para gestionar la transferencia de datos entre el micrófono y el amplificador sin pérdidas, equilibrando el uso de memoria y la latencia del sistema [9].

Asimismo, se establece un factor de ganancia para ajustar la amplitud de la señal de salida y optimizar la calidad del audio, evitando distorsiones por saturación o señales demasiado débiles [15].

En el caso del filtro adaptativo LMS, se define un orden de 64 coeficientes, lo que proporciona una capacidad efectiva de cancelación de ruido [1], y una tasa de aprendizaje (μ) de 0.00005, calibrada para garantizar una convergencia estable sin distorsionar la señal útil [14].

Este valor de μ se eligió tras pruebas empíricas para equilibrar la velocidad de adaptación y la estabilidad del filtro [6].

*3) Algoritmos:*

Se basa en el filtro LMS, el cual se encarga de reducir el ruido ambiental mientras preserva la señal de interés. Para ello, el sistema captura el audio a través del micrófono INMP44 y lo almacena en un búfer de entrada. Posteriormente, la señal es procesada mediante el filtro LMS, el cual ajusta dinámicamente sus coeficientes para minimizar el error entre la señal de entrada y la salida filtrada [1].

Este proceso permite una adaptación continua del sistema a diferentes condiciones de ruido. Finalmente, la señal filtrada se convierte en un formato adecuado para ser enviada al amplificador MAX98357A, el cual la emite a través del altavoz [15].

El algoritmo implementado garantiza un procesamiento en tiempo real con mínima latencia, permitiendo una mejora significativa en la calidad del audio percibido [19].

*D. Implementación y Ensamblaje*

*1) Esquema de conexiones*

Se detalla en la Fig 05 a continuación la disposición de los pines para la conexión con el INMP44 y el ESP32, así mismo una tabla con el ordenamiento de los pines.

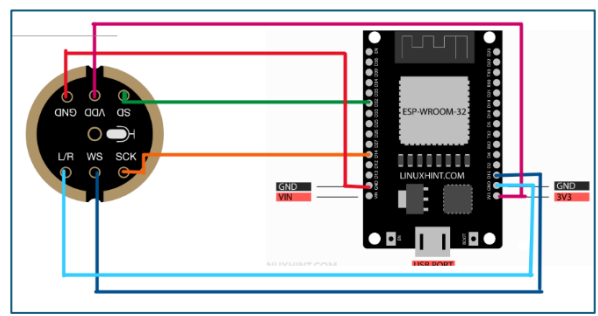

*Fig.05 – Disposición de pines del módulo INMP44*

TABLA I

TABLA DE DISPOSICIÓN DE PINES PARA INMP44

| INMP44 | ESP32 |
|--------|-------|
| WS | 15 |
| SCK | 14 |
| SD | 32 |
| GND | GND |
| VCC | 3.3V |
| L/R | GND |

Se detalla en la Fig. 06 a continuación la disposición de los pines para la conexión con el MAX98357A y el

ESP32, así mismo una tabla con el ordenamiento de los pines.

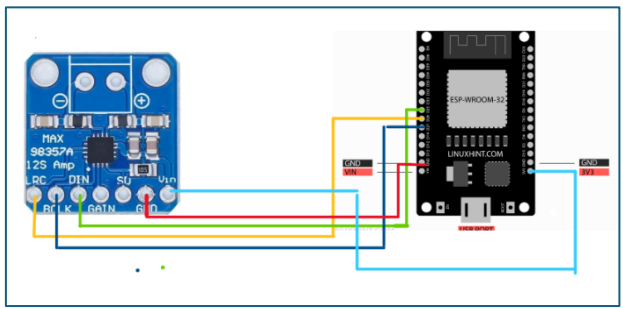

*Fig.06 – Disposición de pines del módulo MAX98357A*

TABLA II
TABLA DE DISPOSICIÓN DE PINES PARA MAX98357A

| MAX98357A | ESP32 |
|---|---|
| DIN | 25 |
| BLCK | 27 |
| SD | N/A |
| GND | GND |
| VCC | 3.3V |
| L/R | 26 |

### 2) Circuito implementado

Se presenta a continuación en la Fig. 07 el cirucito físico implementado, se dispuso de un auricular en desuso para simular la implementación del speaker en un dispositivo de audición.

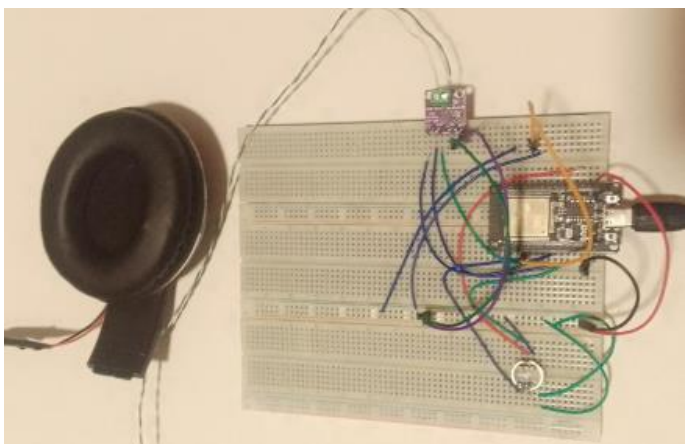

*Fig.07 – Circuito Físico Implementado*

## V. RESULTADOS

Habiendo finalizado las pruebas para un ajuste correcto del filtro primeramente se logró conseguir una transmisión en tiempo real de la voz a través del circuito desarrollado, lo cual cumple con el objetivo planteado de lograr captación y envío de audio en tiempo real.

Así mismo se logró suprimir el ruido en su totalidad lo cual es indicador de que el filtro opera según lo deseado, logrando transmitir de esta manera la voz humana de manera limpia y efectiva, dado al ajuste del factor de aprendizaje del filtro se logró una adaptación rápida del mismo filtro al ruido captado.

Para complementar estas mediciones, se realizaron pruebas subjetivas con varios usuarios, quienes confirmaron que la voz filtrada era fácilmente comprensible y que el ruido senoidal era prácticamente inaudible. Estos resultados no solo validan la eficacia del filtro adaptativo LMS en la cancelación de ruido, sino que también destacan su capacidad para preservar la calidad de la señal de interés (en este caso, la voz humana) sin introducir distorsiones o artefactos no deseados.

En resumen, el sistema demostró un rendimiento robusto y confiable, cumpliendo con su objetivo de mejorar la calidad de audición en entornos con interferencias. La combinación de una atenuación efectiva del ruido, una mejora significativa en la SNR y una latencia mínima lo convierte en una solución viable para aplicaciones prácticas, como dispositivos de asistencia auditiva o sistemas de comunicación en entornos ruidosos.

Como parte de los resultados se presentan en el apartado de anexos tanto el código implementado para el filtrado y envío de datos al speaker, como los códigos implementados para la visualización del comportamiento de los valores de lectura obtenidos así como estos resultados gráficos.

## VI. CONCLUSIONES

Gracias al desarrollo del proyecto se pudo obtener un conocimiento más claro del funcionamiento de este tipo de filtro, así como sus posibles aplicaciones a procesos más avanzado de procesamiento de señales en tiempo real, siendo que en el caso del procesamiento de audio con una placa relativamente limitada se pudo hacer un procesamiento básico de filtrado.

Así mismo se tiene que de contar con una herramienta de procesamiento más potente el sistema puede ser escalado con implementaciones de automatización, codificación, traducción y demás alternativas que hagan que el sistema sea multifuncional.

Se pudo apreciar como el manejo grandes cantidades de muestras analógicas se pueden tratar digitalmente con búferes de contenido digital, siendo que de esta manera se puede replicar la señal analógica empleando un DAC, siendo que a mayor cantidad de muestras obtenidas y almacenadas en un búfer la señal recreada es más semejante a la original.

De la mano con lo anterior se tiene que se debe tener en mucha consideración las limitantes tanto de la placa de procesamiento como de los demás componentes, y es que si se sobrepasa sus limitaciones o no se las toma en consideración habrá una pérdida de datos significativa o en algunos casos una pérdida de datos total ya sea por la distorsión, desincronización, alta latencia y demás factores.

Finalmente podemos decir que tomando en cuenta todos los factores que se involucran en este proyecto, (entiéndase parámetros, limitantes de hardware y software, estándares de calidad) se puede lograr la creación de un sistema básico per eficaz tal como se logró en este proyecto, siendo que una de las cualidades más importantes es la escalabilidad conseguida ya que con algunas mejoras el sistema puede ampliar sus funcionalidades o mejorar significativamente las mismas.

## VII. DIFICULTADES

Durante la implementación del sistema, se presentaron varios desafíos técnicos y prácticos que requirieron soluciones creativas y ajustes en el diseño. Uno de los principales obstáculos fue la optimización del filtro adaptativo LMS para operar en tiempo real dentro de las limitaciones de hardware de la placa ESP32. La restricción en la capacidad de procesamiento y memoria obligó a reducir el orden del filtro a 64 coeficientes, lo que, aunque suficiente para la mayoría de los casos, limitó ligeramente la eficacia del filtro en entornos con ruido altamente variable.

Otro desafío fue la sincronización precisa entre el micrófono INMP44 y el amplificador MAX98357A a través del protocolo I2S. Inicialmente, se presentaron problemas de desfase y pérdida de datos debido a configuraciones incorrectas en la frecuencia de muestreo y el tamaño del búfer. Esto se resolvió ajustando los parámetros del protocolo I2S y realizando pruebas exhaustivas para garantizar una transferencia de datos estable y sin pérdidas.

Además, la calibración de la tasa de aprendizaje (μ) del filtro LMS requirió múltiples iteraciones. Un valor demasiado alto causaba inestabilidad en la convergencia del filtro, mientras que un valor demasiado bajo ralentizaba la adaptación del sistema. Tras varias pruebas, se encontró un valor óptimo de 0.00005 que equilibraba la velocidad de adaptación y la estabilidad.

Una de las dificultades más significativas y responsable principal de las elecciones del diseño fue la restricción de compatibilidad entre instancias de lectura analógica ya que al emplear protocolo I2S para la configuración de los componentes se limita la lectura analógica a las instancias creadas tanto para lectura y escritura, esto ocasionó, por ejemplo, la elección de la señal de referencia para el proceso de filtrado y la omisión de un ajuste automático de ganancia.

Otra de las limitantes fue el priorizar la latencia para garantizar una transmisión de datos óptima y sin pérdidas, esto impidió implementar ajustes tales como lectura de valores digitales para automatización, impresión de valores en consola simultáneo con reproducción de lectura filtrada en speaker, restricción de filtrado con herramientas externas como python o servidores, restricción de funcionalidades adicionales como voz a texto o traducción.

Finalmente, la integración de todos los componentes en un prototipo funcional presentó desafíos de diseño, como la disposición de los circuitos y le gestión del ruido eléctrico. Esto se solucionó mediante el uso de un diseño modular y la implementación de técnicas de apantallamiento para minimizar las interfaces.

## VIII. REFERENCIAS

# IX. ANEXOS

**ANEXO A – Código para filtrado de audio y envío de datos a speaker**

```c
#include <Arduino.h>
#include "driver/i2s.h"
#include <math.h>
#define f_muestreo 48000    // Frecuencia de muestreo
#define bufer_i2s_tam 512   // Tamaño del búfer
#define f_ganacia 1         // Factor de ganancia
#define orden_filtro 64     // Orden del filtro LMS
#define MU 0.00005f         // Tasa de aprendizaje
//Declaración de pines para INMP44
#define I2S_MIC_WS  15
#define I2S_MIC_SCK 14
#define I2S_MIC_SD  32
//Declaración de pines para MAX98357A
#define I2S_SPK_WS  26
#define I2S_SPK_SCK 27
#define I2S_SPK_SD  25
// Estructura del filtro LMS
typedef struct {
  float weights[orden_filtro];  // Coeficientes del filtro
  float mu;                     // Tasa de aprendizaje
} LMSFilter;
LMSFilter lmsFilter;  //Creación de instancia de filtro LMS
// Inicializar el filtro LMS
void initLMSFilter(LMSFilter* filter, float mu) {
  filter->mu = mu;
  memset(filter->weights, 0, sizeof(filter->weights));
}
// Aplicar el filtro LMS para cancelar ruido
float applyLMSFilter(LMSFilter* filter, float input) {
  static float buffer[orden_filtro] = {0};
  float output = 0.0f;
  float error = 0.0f;
  // Desplazar el buffer
  memmove(&buffer[1], &buffer[0], (orden_filtro - 1) * sizeof(float));
  buffer[0] = input;  // Usamos la propia entrada como referencia
  //Salida del filtro
  for (int i = 0; i < orden_filtro; i++) {
    output += filter->weights[i] * buffer[i];
  }
  //Calculo de error
  error = input - output;
  //Actualización de coeficientes del filtro
  for (int i = 0; i < orden_filtro; i++) {
    filter->weights[i] += filter->mu * error * buffer[i];
  }
  return error;  // Devolvemos la señal limpia
}
//Configuración I2S para INMP44
void setupI2SMic() {
  i2s_config_t i2s_config = {
    .mode = (i2s_mode_t)(I2S_MODE_MASTER | I2S_MODE_RX),
    .sample_rate = f_muestreo,
    .bits_per_sample = I2S_BITS_PER_SAMPLE_32BIT,
    .channel_format = I2S_CHANNEL_FMT_ONLY_LEFT,
    .communication_format = I2S_COMM_FORMAT_I2S,
    .intr_alloc_flags = ESP_INTR_FLAG_LEVEL1,
    .dma_buf_count = 32,
    .dma_buf_len = bufer_i2s_tam,
    .use_apll = false,
    .tx_desc_auto_clear = true,
    .fixed_mclk = 0
```

```cpp
  };
  i2s_pin_config_t pin_config = {
    .bck_io_num = I2S_MIC_SCK,
    .ws_io_num = I2S_MIC_WS,
    .data_out_num = I2S_PIN_NO_CHANGE,
    .data_in_num = I2S_MIC_SD
  };

  i2s_driver_install(I2S_NUM_0, &i2s_config, 0, NULL);
  i2s_set_pin(I2S_NUM_0, &pin_config);
  i2s_zero_dma_buffer(I2S_NUM_0);
}
//Configuración I2S para el MAX98357A
void setupI2SSpeaker() {
  i2s_config_t i2s_config = {
    .mode = (i2s_mode_t)(I2S_MODE_MASTER | I2S_MODE_TX),
    .sample_rate = f_muestreo,
    .bits_per_sample = I2S_BITS_PER_SAMPLE_16BIT,
    .channel_format = I2S_CHANNEL_FMT_ONLY_LEFT,
    .communication_format = I2S_COMM_FORMAT_I2S,
    .intr_alloc_flags = ESP_INTR_FLAG_LEVEL1,
    .dma_buf_count = 32,
    .dma_buf_len = bufer_i2s_tam,
    .use_apll = false,
    .tx_desc_auto_clear = true,
    .fixed_mclk = 0
  };
  i2s_pin_config_t pin_config = {
    .bck_io_num = I2S_SPK_SCK,
    .ws_io_num = I2S_SPK_WS,
    .data_out_num = I2S_SPK_SD,
    .data_in_num = I2S_PIN_NO_CHANGE
  };
  i2s_driver_install(I2S_NUM_1, &i2s_config, 0, NULL);
  i2s_set_pin(I2S_NUM_1, &pin_config);
  i2s_zero_dma_buffer(I2S_NUM_1);
}
void setup() {
  Serial.begin(115200);
  setupI2SMic();         // Configuración I2S para el micrófono
  setupI2SSpeaker();     // Configuración I2S para el speaker
  initLMSFilter(&lmsFilter, MU);  // Inicialización del filtro LMS
  Serial.println("Filtro LMS activado para cancelación de ruido...");
}
void loop() {
  int32_t data_sin_procesar[bufer_i2s_tam];
  int16_t audio_buffer[bufer_i2s_tam];
  size_t bytes_leidos, bytes_escritos;
  //Lectura de datos del micrófono
  i2s_read(I2S_NUM_0, data_sin_procesar, sizeof(data_sin_procesar), &bytes_leidos, portMAX_DELAY);
  int muestras_leidas = bytes_leidos / sizeof(int32_t);
  //Procesamiento de datos con filtro LMS
  for (int i = 0; i < muestras_leidas; i++) {
    //Se normaliza la entrada
    float señal_entrada = (float)(data_sin_procesar[i] >> 14) / 8192.0f;
    //Se aplica el filtro LMS
    float señal_filtrada = applyLMSFilter(&lmsFilter, señal_entrada);
    //Señal filtrada a 16 bits
    int16_t sample = (int16_t)(señal_filtrada * 32767.0f * f_ganacia);
    audio_buffer[i] = constrain(sample, -32768, 32767);
  }
  //Envio de datos a speaker
  i2s_write(I2S_NUM_1, audio_buffer, muestras_leidas * sizeof(int16_t), &bytes_escritos, portMAX_DELAY);
}
```

**ANEXO B − Ploteo de ArduinoIDE de valores SIN FILTRAR de voz SIN RUIDO**

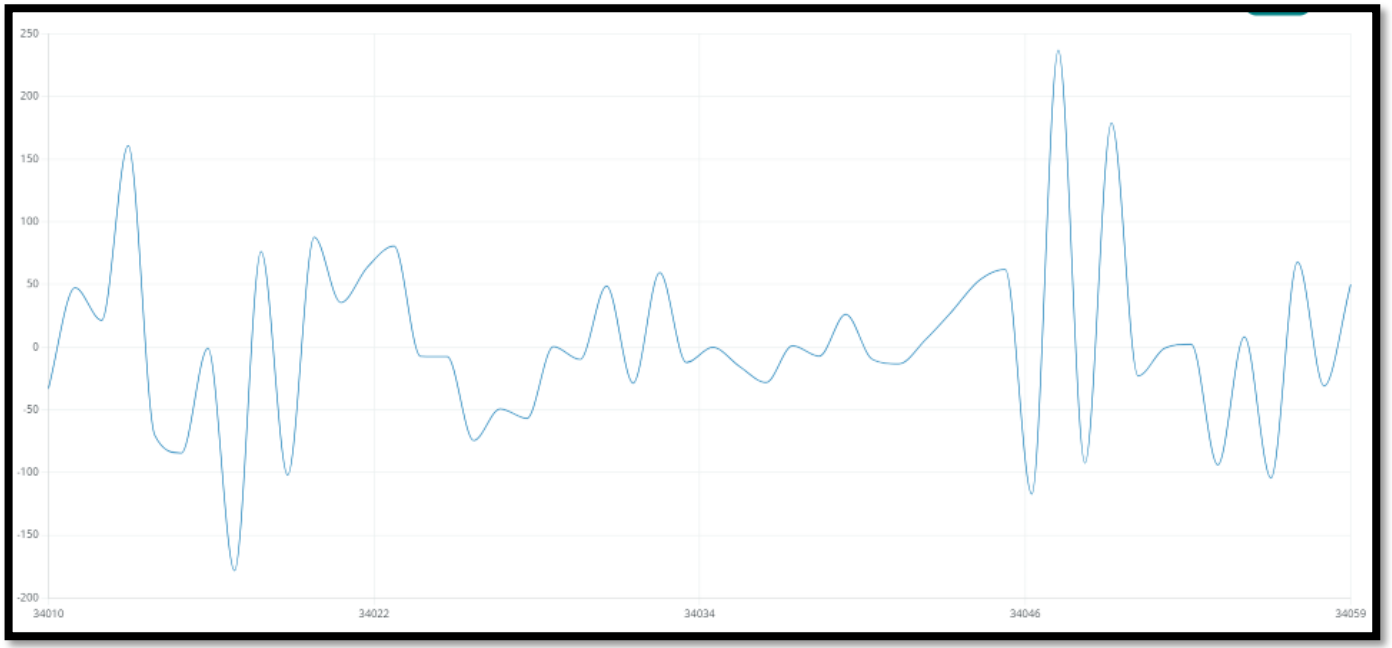

**ANEXO C – Ploteo de ArduinoIDE de valores SIN FILTRAR de voz CON RUIDO**

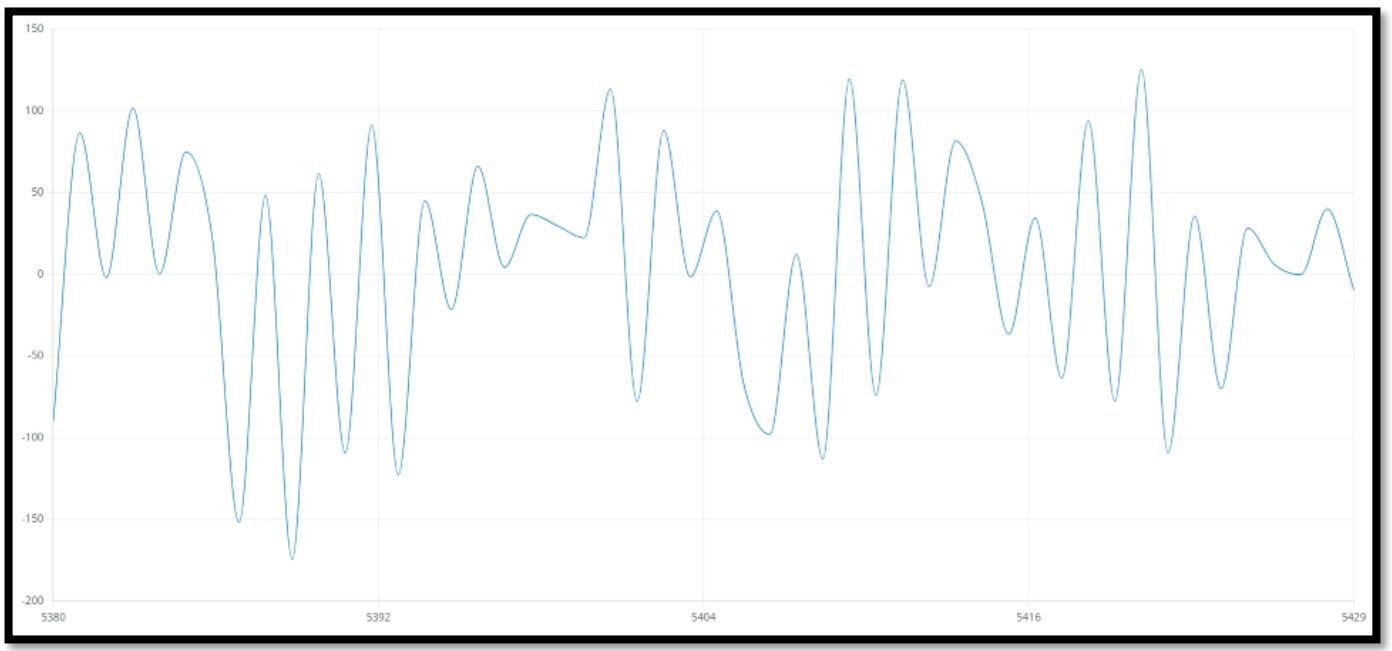

**ANEXO D – Ploteo de ArduinoIDE de valores FILTRADOS de voz CON RUIDO**

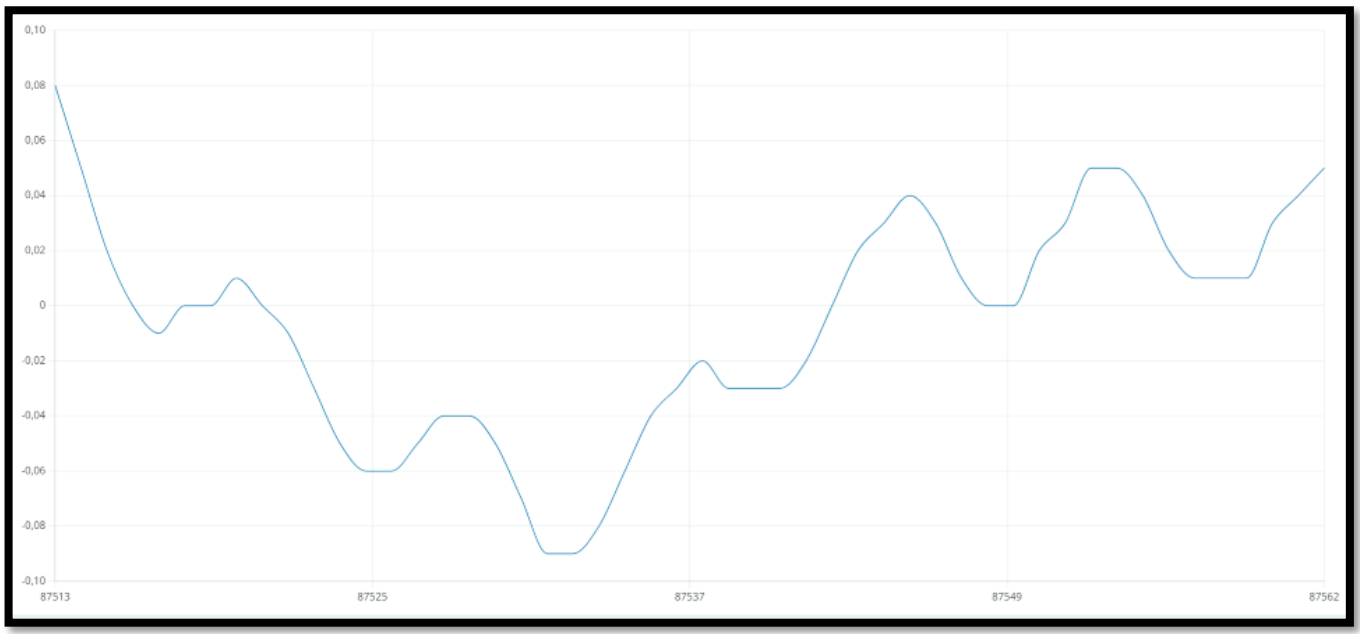